\documentclass{article}
\usepackage[T1]{fontenc} 
\usepackage[utf8]{inputenc} 
\usepackage{ismir,amsmath,cite,url}
\usepackage{graphicx}
\usepackage{color}
\usepackage{amsmath,amssymb,multirow, booktabs, subcaption}

\title{Semi-Supervised Music Tagging Transformer}

 \multauthor
{
	Minz Won$^{1, 2}$  \hspace{.7cm} 
	Keunwoo Choi$^2$         \hspace{.7cm}
	Xavier Serra$^1$
} 
{
	$^1$ Music Technology Group, Universitat Pompeu Fabra, Barcelona, Spain\\
	$^2$ ByteDance, Mountain View, California, United States\\
	{\tt\small minz.won@upf.edu}
}
\def\authorname{Minz Won, Keunwoo Choi, and Xavier Serra}

\begin{document}

\maketitle
\begin{abstract}
We present Music Tagging Transformer that is trained with a semi-supervised approach. The proposed model captures local acoustic characteristics in shallow convolutional layers, then temporally summarizes the sequence of the extracted features using stacked self-attention layers. Through a careful model assessment, we first show that the proposed architecture outperforms  the previous state-of-the-art music tagging models that are based on convolutional neural networks under a supervised scheme.

The Music Tagging Transformer is further improved by noisy student training, a semi-supervised approach that leverages both labeled and unlabeled data combined with data augmentation. To our best knowledge, this is the first attempt to utilize the entire audio of the million song dataset. 
\end{abstract}
%

\section{Introduction}\label{sec:introduction}
Automatic music tagging is a classification task whose objective is computational understanding of music semantics. 
From a given audio excerpt, a trained music tagging model predicts relevant tags (e.g., genre, mood, instrument, decade, region) based on its acoustic characteristics. 
The task has attracted music information retrieval (MIR) researchers due to its wide pragmatic usages in many applications. Especially, there is a strong demand from industries that have music recommendation services from large-scale music libraries. 
Thanks to the recent advances in deep learning, mostly convolutional neural networks (CNNs)~\cite{krizhevsky2012imagenet}, the performances of music tagging models have been significantly enhanced by leveraging large-scale data with various deep architectures~\cite{choi2016automatic,pons2018end,lee2017sample,won2020data}. However, there still are two limitations in the current music tagging research: i) chunk-based prediction and ii) a limited amount of labeled data for supervised learning. 

Music signals are in the form of sequential data. In this sequence, regarding typical tags,
some acoustic characteristics may appear locally (e.g., instruments) 
while some others may span over the sequence (e.g., mood, genre). 
This means a successful music tagging model needs to be able to extract both local and global features.
Fully convolutional network~\cite{choi2016automatic}, one of the very early deep learning models for music tagging, was designed to capture 
both local and global features by increasing the size of the overall receptive fields with max-pooling.
More recently, however, it is shown that training with a smaller hop size with shorter audio chunks is beneficial for music tagging~\cite{won2020evaluation}. This approach has been adopted in many CNN-based models~\cite{pons2018end,lee2017sample,won2020data}, where the models are trained with short audio chunks (3~to~5~second long), densely striding max-pooling, and a global pooling layer.
To predict music tags of a 3-minute song, for example, the audio is split into multiple short audio chunks, and the model makes predictions on each chunk. Then, the predictions are aggregated through majority vote or global average-/max-pooling. This means that on a track level, the current music tagging models are performing like a bag-of-features model~\cite{brendel2019approximating} instead of modeling music representation as a sequence. 


Another limitation of the current music tagging research is a limited amount of labeled data. The Modern deep learning models are data-hungry. 
However, manually labeling music tags is time-consuming and requires domain expertise.
In pursuit of large-scale research, the million song dataset (MSD)~\cite{bertin2011million}, which literally includes a million songs in it, became popular in music tagging research.
Among the million songs, however, only about 24\% are labeled with at least one of the top-50 music tags. Most of the previous music tagging research has only utilized the labeled data while discarding 76\% of the songs in the dataset. This type of setup (i.e., a small labeled dataset along with a large unlabeled dataset) is not limited to the MSD but can be found often in the real world regardless of the domain. To leverage the unlabeled data, self-supervised~\cite{oord2018representation,he2020momentum,chen2020simple,grill2020bootstrap,spijkervet2021contrastive} and semi-supervised~\cite{chapelle2009semi,xie2020self,chen2020big} learning have been actively explored in computer vision and natural language processing.

In this paper, we present Music Tagging Transformer that is trained with a semi-supervised approach.
Our main contribution is three-fold: (\textit{i})
through a careful model assessment, we show that our Music Tagging Transformer outperforms the previous works, (\textit{ii}) we show that we can use unlabeled data to improve music tagging performances via semi-supervised learning, (\textit{iii}) we provide a new split of the MSD to solve known issues of the previous one.
%
Reproducible code is available online.~\footnote{https://github.com/minzwon/semi-supervised-music-tagging-transformer}

\section{Related Works}\label{sec:related}
\subsection{Music Tagging}
\subsubsection{CNN-based models}

The research using CNNs accelerated the improvement of music tagging models. Choi~et~al. proposed to adopt CNNs for music tagging~\cite{choi2016automatic}. Pons~et~al. intended to inject domain knowledge in their architecture by designing vertically and horizontally long filters so that the network can capture timbral and temporal features, respectively~\cite{pons2018end}. Lee~et~al. formalized music tagging as a fully end-to-end process by using a 1D CNN and raw audio inputs~\cite{lee2017sample}. Won~et~al. proposed to use a learnable harmonic front end for a 2D CNN~\cite{won2020data}. More recently, the authors of \cite{won2020evaluation} evaluated all these CNN-based music tagging models under the same experimental environment.
Two of the main conclusions of \cite{won2020evaluation} are to recommend (\textit{i}) using mel spectrogram inputs, and (\textit{ii}) using the most granular 2D filters (i.e., $3 \times 3$ convolution) instead of manual design choices. That means, a simple 2D CNN with mel spectrogram inputs, which is prevalent and sometimes referred to as \textit{vgg-ish} model, is still outperforming the other music tagging models.

\subsubsection{Multiple instance learning}
Music tagging can be seen as a multiple instance learning (MIL)~\cite{dietterich1997solving} problem. 
A given music signal (a bag of multiple instances) will be labeled with a tag if a part (an instance) of the signal has a certain, relevant acoustic characteristic. The acoustic characteristic may span across the entire sequence (e.g., mood) or appear partially (e.g., instruments). There are two approaches of handling this aspect in music tagging. One is to train an instance-level model then aggregate the instance-level predictions; the other is to handle multiple instances in a single model.

The first approach has been used in many systems~\cite{pons2018end,lee2017sample,won2020data}.
This approach is justified by our intuition -- humans can predict music tags within just a few seconds. For example, people would not spend 3~minutes to determine whether a track is \textit{rock}. 
In this approach, during the evaluation phase, the instance-level predictions are aggregated with a method such as majority vote, global max pooling, global average pooling, or adaptive pooling~\cite{mcfee2018adaptive}. 

The second approach, the single model one, tackles the MIL problem in an end-to-end fashion. Fully convolutional network~\cite{choi2016automatic}, an early music tagging model using CNN, models the entire 30-second mel spectrogram inputs. 
However, this model has to stick to a fixed input size and shows a relatively lower performance compared to other models that are trained with short audio chunks. 
To take the global structure into account while not loosing local features, sequential modeling was added to the deep learning architectures. 
Convolutional recurrent neural network (CRNN)~\cite{choi2017convolutional} is designed to capture local acoustic characteristics in a CNN front end and to summarize the sequence of the extracted features using an RNN back end. Similarly, another previous work~\cite{won2019toward} adopted a sequence model from the natural language processing field, the Transformer~\cite{devlin2018bert}.

\subsection{Transformers}
\subsubsection{Overview}
Transformer~\cite{vaswani2017attention,devlin2018bert} has shown its ability in sequence modeling, establishing itself  \textit{de facto} state-of-the-art in natural language processing. The structure of Transformer is a deep stack of self-attention layers. In each layer, by self-attention mechanism, the pairwise attention scores between every time step are calculated to output another sequence which includes a better context in each time step. In detail, the self-attention score is calculated as:
\begin{equation}\label{eqn:attention}
Attention(Q, K, V) = softmax\left(\frac{QK^T}{\sqrt{d_k}}\right)V,
\end{equation}
where $d_k$ is the dimension of the \textit{key} and $Q$, $K$, and $V$ are \textit{query}, \textit{key}, and \textit{value}, respectively. Based on the relevance of \textit{key} and \textit{query} (i.e., relationship between two items), how much of \textit{value} to be passed to the next layer is decided. Since Transformer has direct paths between each time step, it does not suffer from \textit{vanishing gradient}, which is a critical problem of RNN families when modeling long sequences. However, Transformer may suffer from its memory complexity which is quadratic to the length of the sequence.

Due to its receptive field, which is unlimited within the input sequence, Transformer has become extremely popular in sequential data modeling such as text~\cite{devlin2018bert}, symbolic music~\cite{huang2018music}, and video~\cite{arnab2021vivit}. It even demonstrated its representation power in non-sequential data such as image~\cite{dosovitskiy2020image}. 


\subsubsection{Self-attention in music tagging}
Hereinafter, we refer to the previous work~\cite{won2019toward} as CNNSA (acronym of convolutional neural network with self-attention) so that we can distinguish it from the proposed Music Tagging Transformer. The CNNSA model is the first music tagging model that uses Transformer. The model consists of a CNN front end and a Transformer back end. The CNN front end captures local (0.03 to 2.6 seconds) acoustic characteristics and the Transformer back end temporally summarizes the sequence of the features. Based on previous works, the CNNSA investigated two types of CNN front end which have two very different motivations. One is based on hand-crafted vertically and horizontally long filters~\cite{pons2018end} to capture timbral and temporal characteristics, respectively. The other is a more flexible, data-driven approach by using 1D CNN~\cite{lee2017sample} with raw audio inputs.
The CNNSA showed comparable results in music tagging but did not outperform other previous CNN approaches. Nevertheless, it demonstrated the Transformer's sequence modeling ability and better temporal interpretability.

\subsection{Semi-supervised Learning}
With the advances of scalable hardware and training algorithms, the demand for labeled data has outpaced the progress of the size of datasets in many fields. 
As a solution, researchers started to develop methods that can take advantage of unlabeled data. 
Self-supervised~\cite{oord2018representation,he2020momentum,chen2020simple,grill2020bootstrap} and semi-supervised learning~\cite{chapelle2009semi,xie2020self,chen2020big} aim at leveraging the abundant unlabeled data and have shown strong performances in various domains including computer vision and natural language processing. 

In many self- and semi-supervised learning approaches, the models are trained to return noise-invariant predictions~\cite{chen2020simple,grill2020bootstrap,xie2020self}. When there is an apple on the table, for example, it is always an apple even if we take a look at it from a different angle or a different distance, under different lights, or through glass. 
This `noise' is usually realized in a form of data augmentation.
In detail, with a self-supervised learning scheme, models are trained to optimize the agreement between different views of the same input~\cite{chen2020simple,grill2020bootstrap}. 

On the contrary, semi-supervised learning takes advantage of \textit{both} existing labeled data and unlabeled data. One effective way of handling the data is to formalize the problem as teacher-student learning~\cite{li2017large,xie2020self,kum2020semi}. In teacher-student learning, a teacher model is first trained with labeled data in a supervised scheme and then, a student model is trained to mimic the teacher's behavior by predicting pseudo-labels~\cite{lee2013pseudo} that are generated by the teacher model. The teacher-student training has been actively explored with the purpose of domain adaptation~\cite{li2017large}, knowledge distillation~\cite{kim2016sequence}, and knowledge expansion~\cite{xie2020self}. Especially, noisy student training~\cite{xie2020self} successfully takes advantage of the teacher-student training with the aforementioned noise invariance.

\section{Models}\label{sec:models}
\subsection{Short-chunk ResNet}
As a simple 2D CNN with mel spectrogram inputs outperforms other music tagging models~\cite{won2020evaluation}, we use a short-chunk ResNet model as our baseline. The model is trained with 3.69-second short audio chunks (instance-level), then the predictions are later aggregated by averaging them in the evaluation phase. It is a seven-layer CNN and each layer comprises of $3 \times 3$ convolution with residual connection\cite{he2016deep}, batch normalization~\cite{ioffe2015batch}, rectified linear unit (ReLU) non linearity, and $2 \times 2$ max pooling. This is a variant of the prevalent \textit{vgg-ish} model, but due to its specific characteristics (i.e., short-instance-level training and densely striding max pooling), it is referred to as short-chunk ResNet~\cite{won2020evaluation}. We used a publicly available implementation~\footnote{https://github.com/minzwon/sota-music-tagging-models}.



\subsection{Music Tagging Transformer}
\begin{figure}
 \centerline{
 \includegraphics[width=0.8\columnwidth, trim=4 0 0 0, clip]{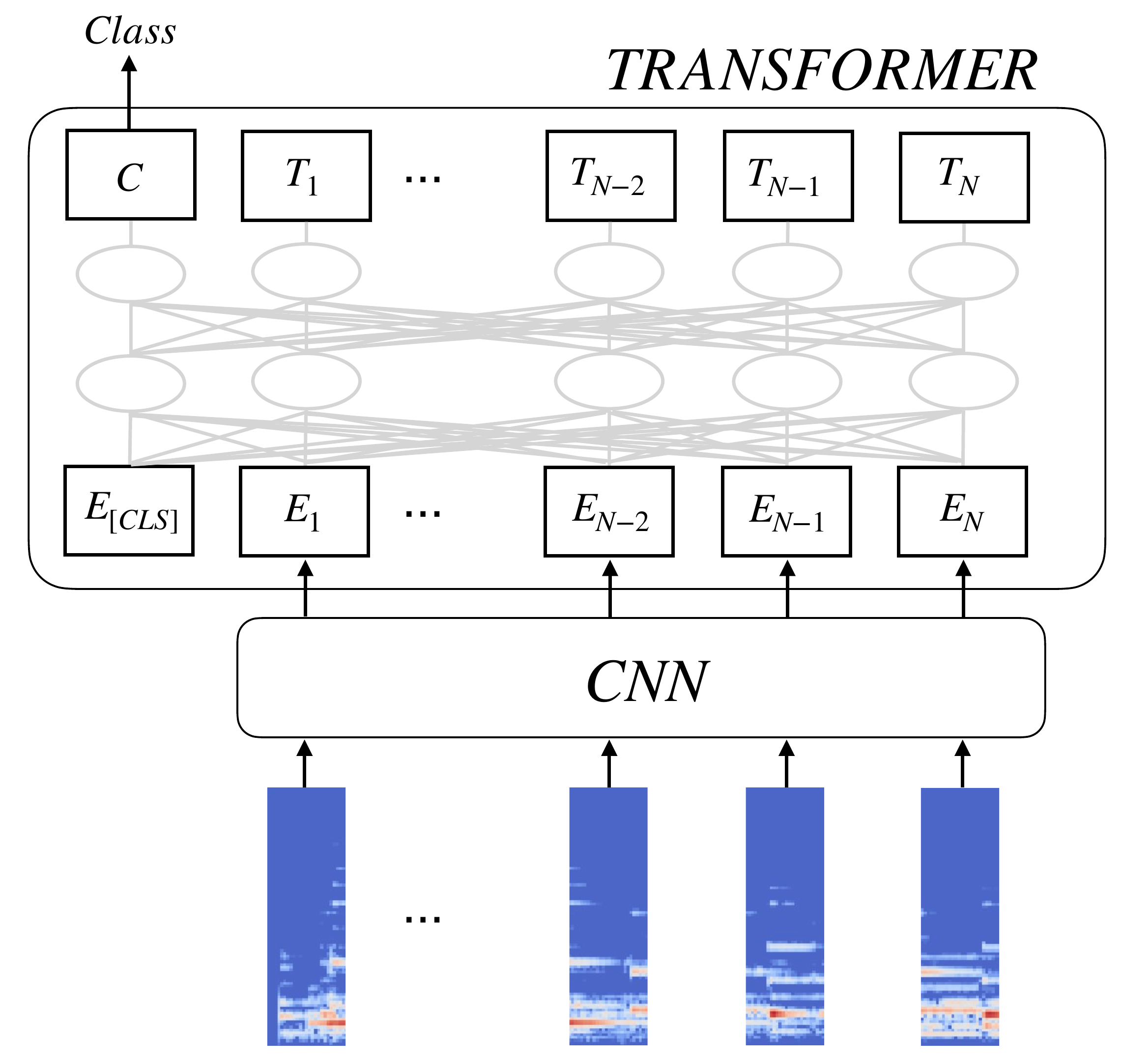}}
 \caption{Proposed Music Tagging Transformer.}
 \label{fig:model}
\end{figure}

\begin{table}[t!]
\centering
\begin{tabular}{@{}lc@{}}
\toprule
layer & output shape \\ \midrule
Input & $B \times 1 \times F \times T$ \\
Conv ($3\times3$) & $B \times C \times F \times T$ \\
MaxPool ($2\times2$) & $B \times C \times F/2 \times T/2$ \\
Conv ($3\times3$) & $B \times C \times F/2 \times T/2$ \\
MaxPool ($2\times2$) & $B \times C \times F/4 \times T/4$ \\
Conv ($3\times3$) & $B \times C \times F/4 \times T/4$ \\
MaxPool ($2\times1$) & $B \times C \times F/8 \times T/4$ \\
Reshape & $B \times (C \cdot F/8) \times T/4$ \\
Fully-connected & $B \times C^{\prime} \times T/4$ \\

\bottomrule
\end{tabular}
\caption{Front end CNN of Music Tagging Transformer.}
\label{tab:frontend}
\end{table}

The proposed Music Tagging Transformer consists of two parts: a CNN front end and a Transformer back end as in Figure~\ref{fig:model}. On a high level, our structure is similar to the previous CNNSA~\cite{won2019toward} where CNN captures local spectro-temporal features and Transformer globally summarizes the sequence of the extracted features. However, there are two main differences in the front end: the CNN architecture and the reshaping layer. 

The CNNSA~\cite{won2019toward} investigated two different CNN front ends that are from two opposite motivations. One is using hand-crafted filter design on mel spectrogram inputs to leverage domain knowledge~\cite{pons2018end}, the other is, with a fully data-driven spirit, one that learns from raw audio in an end-to-end fashion~\cite{lee2017sample}. However, we follow a suggestion from a more recent work~\cite{won2020evaluation} 
--
we use $3\times3$ convolution filters with residual connections~\cite{he2016deep} on mel spectrogram inputs. Table~\ref{tab:frontend} outlines our 3-layered CNN front end where $B$ is the batch size, $C$ is the number of convolution channels, $F$ is the number of mel bins, $T$ is the number of frames, and $C^{\prime}$ is the number of attention channels of Transformer. 
This CNN front end i) helps the model to capture local representations and ii) reduces the time resolution of the input so that it is feasible to train the following back end.

At the end of the CNN, the second and the third dimensions are reshaped into a single dimension. This flattening is motivated by Vision Transformer (ViT)~\cite{dosovitskiy2020image} which reshapes a 2D image patch into a one-dimensional array. 
As a result, the output of the CNN is a sequence of short-chunk audio features where a chunk corresponds to approximately 0.1~second. It is input to the back end Transformer.
This is in contrast to the CNNSA~\cite{won2019toward} that used the frequency-axis max-pooling at the end of its front end. In other words, in Music Tagging Transformer, the attention layers are given more detailed spectral information. 


Our back end Transformer architecture is nearly identical to the previous works~\cite{devlin2018bert,won2019toward} except for the number of parameters and input lengths. 
After a hyperparameter search, we chose 4~layers, 256~attention dimensions, and 8~attention heads. At the input stage, positional embedding~\cite{devlin2018bert} is applied and a special token embedding $E_{[\textit{CLS}]}$ is inserted so that the Transformer can perform sequence classification (Figure~\ref{fig:model}) as a downstream task.



\subsection{Noisy Student Training}
To leverage unlabeled data, we investigate noisy student training~\cite{xie2020self}, a successful semi-supervised learning approach. Table~\ref{tab:pseudocode} provides an overview of noisy student training with pseudocode. First, we train a teacher model $\mathcal{T}$ with a conventional supervised learning approach (line 1--6). Then, we train a student model $\mathcal{S}$ with two types of losses. The first loss, $l_1$, is coming from the typical supervised approach with labeled data (line 10--11) as done for the teacher model. The other loss, $l_2$, is from unlabeled inputs and the corresponding pseudo-labels provided by the teacher model (line 12--15). In order to make the student model perform beyond mimicking the teacher model, data augmentation is applied (line 13). Both hard and soft labels can be used for the pseudo-labels~\cite{xie2020self} and we use soft labels in our work. If the training is successful, the trained student model would outperform the teacher model. Furthermore, the whole training process can be done iteratively by using the student as a new teacher model and training another student model to obtain an even better performing model. 
For a stronger teacher model, we used data augmentation in our supervised learning pipeline as well (line 1--6 and 10--11). As a result, the only pipeline without data augmentation is pseudo-label generation (line 12).

The size of the student model can be identical or larger than the teacher model. In this case, one can interpret the training process as \textit{knowledge expansion}~\cite{xie2020self}, meaning the knowledge in the teacher model is upgraded in the student model. One can also design the student model to be smaller than the teacher model, making the process \textit{knowledge distillation}~\cite{hinton2015distilling}. Knowledge expansion and knowledge distillation are complementary; depending on the use-case, one could pursue either performance or efficiency. We investigate both directions in this paper.

\begin{table}[t!]
\centering
\begin{tabular}{@{}l@{}}
\toprule
\textbf{Noisy Student Training} \\ \midrule
\textbf{Input} labeled data $X$, labels $Y$, unlabeled data $Z$ \\
\textbf{Models} teacher model $\mathcal{T}$, student model $\mathcal{S}$ \\
\textbf{Functions} loss function $\mathcal{L}$, data augmentation $\mathcal{A}$, \\
\hspace{16mm}back propagation $\mathcal{B}$ \\
\textbf{Train} \\
1\hspace{5mm}\textbf{for} $x \in X$, $y \in Y$ \\
2\hspace{5mm}\hspace{5mm}\textbf{do} \\
3\hspace{5mm}\hspace{5mm}$p$ $\longleftarrow$ $\mathcal{T}(x)$ \hspace{11.45mm}// predict\\
4\hspace{5mm}\hspace{5mm} $l$ $\longleftarrow$ $\mathcal{L}(p, y)$\hspace{9.64mm}// get loss\\
5\hspace{5mm}\hspace{5mm}$\mathcal{T}\longleftarrow\mathcal{B}(\mathcal{T}, l)$ \hspace{7.75mm}// update teacher model\\
6\hspace{5mm}\hspace{5mm}\textbf{end do} \\
7\hspace{5mm}\textbf{end for} \\
8\hspace{5mm}\textbf{for} $x \in X$, $y \in Y$, $z \in Z$\\
9\hspace{5mm}\hspace{5mm}\textbf{do} \\
10\hspace{3.5mm}\hspace{5mm} $p_1$ $\longleftarrow$ $\mathcal{S}(x)$\hspace{10.38mm}// predict\\
11\hspace{3.5mm}\hspace{5mm} $l_1$ $\longleftarrow$ $\mathcal{L}(p_1, y)$\hspace{6.75mm}// get supervised loss\\
12\hspace{3.5mm}\hspace{5mm} $\psi$ $\longleftarrow$ $\mathcal{T}(z)$\hspace{11.15mm}// generate pseudo-label\\
13\hspace{3.5mm}\hspace{5mm} $\hat{z}$ $\longleftarrow$ $\mathcal{A}(z)$\hspace{11.73mm}// data augmentation\\
14\hspace{3.5mm}\hspace{5mm} $p_2$ $\longleftarrow$ $\mathcal{S}(\hat{z})$\hspace{10.65mm}// predict\\
15\hspace{3.5mm}\hspace{5mm} $l_2$ $\longleftarrow$ $\mathcal{L}(p_2, \psi)$\hspace{6.4mm}// get semi-supervised loss\\
16\hspace{3.5mm}\hspace{5mm}$\mathcal{S}\longleftarrow\mathcal{B}(\mathcal{S}, l_1+l_2)$ \hspace{1.2mm}// update student model\\
17\hspace{3.5mm}\hspace{5mm}\textbf{end do} \\
18\hspace{3.5mm}\textbf{end for} \\

\bottomrule
\end{tabular}
\caption{Pseudocode of noisy student training.}
\label{tab:pseudocode}
\end{table}

\subsection{Data Augmentation}
Data augmentation is a key to success in noisy student training. In our experiments, we take advantage of \textit{Audio Augmentations} library~\cite{spijkervet_torchaudio_augmentations} which is easily integrated to PyTorch data pipeline. The applied data augmentation methods are as follows:

\begin{itemize}
    \item \noindent\textbf{Polarity inversion}.
    \item \noindent\textbf{Additive noise} by $k_{snr} \in \{0.3, 0.5\}$.
    \item \noindent\textbf{Random gain} by $\mathcal{A} \in \{-20, -1\}$ dB.
    \item \noindent\textbf{High-pass filter} by $f_{H} \in \{2200, 4000\}$ Hz.
    \item \noindent\textbf{Low-pass filter} by $f_{L} \in \{200, 1200\}$ Hz.
    \item \noindent\textbf{Delay} by $t \in \{200, 500\}$ ms.
    \item \noindent\textbf{Pitch shift} by $n \in \{-7, 7\}$ semitones.
    \item \noindent\textbf{Reverb} by room size $s \in \{0, 100\}$.
\end{itemize}
Each augmentation method is activated independently with a probability $p \in \{0.3, 0.7\}$.



\section{Dataset}\label{sec:dataset}
We use the million song dataset (MSD)~\cite{bertin2011million} which consists of one million songs with audio features and metadata. Most of the previous works~\cite{choi2016automatic,pons2018end,lee2017sample,won2020data,won2020evaluation} relied on the Last.fm tags, a set of crowdsourced music tags, as ground truth. A popular approach is to take the most frequent 50~tags and select tracks that have at least one of the tags. This results in 242k songs, which are split into train, validation, and test sets.~\footnote{{https://github.com/keunwoochoi/MSD\_split\_for\_tagging}} Hereinafter, we refer to this as a conventional (MSD) split. Since this split has been widely used, we use it to benchmark the proposed Music Tagging Transformer against the previous works.

We also suggest a new split to alleviate some known problems of the conventional split. There are two problems -- First, since the MSD music tags are collected from users, some of them are very noisy, and that may lead to noisy (and incorrect) evaluation~\cite{choi2017effects}. 
Second, a strict split of music items requires taking the artist information into consideration since often, songs and labels from the same artist heavily resemble each other. 
However, the conventional split was done without such consideration, having caused unintended information leakage between the training and evaluation sets. Ultimately, this would cause an overly optimistic evaluation. 
As a solution, we use manually cleaned data from a previous work~\cite{won2020multimodal} and take the top 50 tags. 
We also propose a new split of MSD that does not share any artist among training/validation/test sets and is extended to more tracks. We name this `CALS split' (cleaned and artist-level stratified split). CALS split consists of 233k labeled tracks and 516k unlabeled tracks. To the best of our knowledge, this is the first attempt to utilize the entire MSD audio, although we have to discard the rest 250k tracks to avoid information leakage by shared artists.

When noisy student training is used, it is common to have an unlabeled set that is significantly bigger than the labeled set. For example, in computer vision, 81~million unlabeled items were used along with 1.2~million labeled items~\cite{xie2020self}, making the ratio of the semi-supervised set to be 67.5 ($81 / 1.2$). 
However, with 233k labeled items and 516k unlabeled items, our ratio is only around 2.3.
This might be a factor that limits us from fully exploring the potential advantage of semi-supervised learning as we will discuss in Section~\ref{sec:results}. 




\section{Experiment}\label{sec:results}

We use the MSD that was introduced in detail in Section~\ref{sec:dataset}. All the audio signals are pre-processed to 22,050~Hz sample rate and converted to short-time Fourier transform representations with a 1024-point FFT and 50\%-overlapping Hann window. Finally, we convert them to log mel spectrograms with 128 mel bins.

All models that are used or introduced in this paper are optimized using Adam~\cite{kingma2014adam} with a learning rate of 0.0001. The best model is selected based on the binary cross entropy loss of the validation set and early stopping is applied when the validation loss does not improve for 20 epochs.

Music tagging models are typically evaluated with Area Under Receiver Operating Characteristic Curve (ROC-AUC). However, it is known that ROC-AUC may report overly optimistic results with highly skewed data~\cite{davis2006relationship}. Therefore, as our main evaluation metrics, we use not only ROC-AUC but also Area Under Precision-Recall Curve (PR-AUC).

\subsection{Performance with the conventional split}
Table~\ref{tab:performance} summarizes the performance of previous systems and the proposed model using the conventional split. The ROC-AUC and PR-AUC of the many previous models have been under 0.89 and 0.33, respectively. Our model, Music Tagging Transformer, outperforms the previous state-of-the-art models, harmonic CNN and short-chunk ResNet. The improvement, especially on PR-AUC, is non-trivial and even larger with data augmentation. 
%

The front end of our Music Tagging Transformer takes a sequence of chunks, where each of which represents a very short duration of the signal ($\approx$0.1 second)
~\cite{won2020evaluation}. 
Because 0.1~second would be too short to represent musical characteristics alone, we interpret that the experimental results would mean our Transformer back end plays a role of sequential feature extractor beyond simple bag-of-feature aggregation. This may be an important aspect of the proposed model since sequential modeling is what the self-attention mechanism is the best suit. 

The data augmentation we adopted contributes to improvements of 0.0056 ROC-AUC and 0.0119 PR-AUC. These are bigger than many of the improvements we have seen between different architecture choices. This emphasizes that data augmentation should be considered when developing a music tagging model.

\begin{table}[t!]
\centering
\begin{tabular}{@{}lcc@{}}
\toprule
Models & ROC-AUC & PR-AUC \\ \midrule
FCN~\cite{choi2016automatic} & 0.8742 & 0.2963  \\
Musicnn~\cite{pons2018end} & 0.8788 & 0.3036  \\
Sample-level~\cite{lee2017sample} & 0.8789    & 0.2959    \\
Sample-level+SE~\cite{kim2018sample} & 0.8838    & 0.3109    \\
CRNN~\cite{choi2017convolutional} & 0.8460    & 0.2330    \\
CNNSA~\cite{won2019toward} & 0.8810    & 0.3103    \\
Harmonic CNN~\cite{won2020data} & 0.8898    & 0.3298    \\
Short-chunk CNN~\cite{won2020evaluation} & 0.8883    & 0.3251    \\
Short-chunk ResNet~\cite{won2020evaluation} & 0.8898    & 0.3280    \\ \midrule
 Transformer (proposed)$^{\mathsection}$   & \textbf{0.8916}  & \textbf{0.3358} \\
Transformer (proposed) + DA$^{\dagger}$  & \textbf{0.8972}  & \textbf{0.3479} \\
\bottomrule
\end{tabular}
\caption{Performance comparison using the conventional MSD split for top-50 music tagging. The ${\mathsection}$ and $\dagger$ marks mean they are based on the identical model architecture and training strategy; compared to the same, marked models in Table~\ref{tab:semi}, only the dataset split is different.}
\label{tab:performance}
\end{table}


\begin{table}[t!]
\centering
\begin{tabular}{@{}lccc@{}}
\toprule
Models & \#param & ROC-AUC & PR-AUC \\ \midrule
ResNet~\cite{won2020evaluation} & 13.5m & 0.9098    & 0.3525    \\
ResNet+DA & 13.5m & 0.9141    & 0.3705    \\
ResNet+DA+KE & 13.5m & 0.9165    & 0.3728    \\ 
ResNet+DA+KD & 3.4m & \textbf{0.9171}    & \textbf{0.3742}    \\\midrule
Transformer$^{\mathsection}$   & 4.6m & 0.9188  & 0.3775 \\
Transformer+DA$^{\dagger}$   & 4.6m & 0.9191  & 0.3845 \\
Transformer+DA+KE   & 4.6m & 0.9204  & 0.3839 \\
Transformer+DA+KD   & 0.5m & \textbf{0.9217}  & \textbf{0.3889} \\ 


\bottomrule
\end{tabular}
\caption{Performance comparison using the CALS MSD split for music tagging.}
\label{tab:semi}
\end{table}



\subsection{Performance with the CALS split}

For a deeper and more accurate analysis of the proposed models and methods, we use the proposed CALS split and run experiments with various configurations. Table~\ref{tab:semi} presents the experimental results of short-chunk ResNets (baseline architecture) and Music Tagging Transformers. For both of the models, it also summarizes the results of supervised models (the baseline among training methods), models with data augmentation (DA), models with DA and knowledge expansion (KE), and models with DA and knowledge distillation (KD). Note that the two bottom rows of Table~\ref{tab:performance} correspond to the 5th and 6th rows of Table~\ref{tab:semi} as marked with ${\mathsection}$ and $\dagger$.

For both short-chunk ResNet~\cite{won2020evaluation} and the Music Tagging Transformer, we observe constant improvements when data augmentation and noisy student training (knowledge expansion) are applied accumulatively. This shows that for both of the architectures, the size of the dataset is a factor that limits the performance of the models.

In Section~\ref{sec:dataset}, we mentioned that our ratio of the semi-supervised set is relatively small. There are two observations that may be related to it.
First, unlike a previous work in computer vision~\cite{xie2020self}, we could not observe any performance gain by iterating the noisy student training (i.e., repeating to use a student model as the next teacher model). 
Second, interestingly, the student model with smaller parameters (models with DA and KD) showed better performance than larger models (models with DA and KE). 
%
This would be explained more clearly if the models are trained with a significantly richer dataset, one that is bigger and/or has more diverse data. Unfortunately, we could not run such an experiment due to the lack of a suitable dataset. 

\begin{figure}
 \centerline{
 \includegraphics[width=1.0\columnwidth, trim=0 0 0 0, clip]{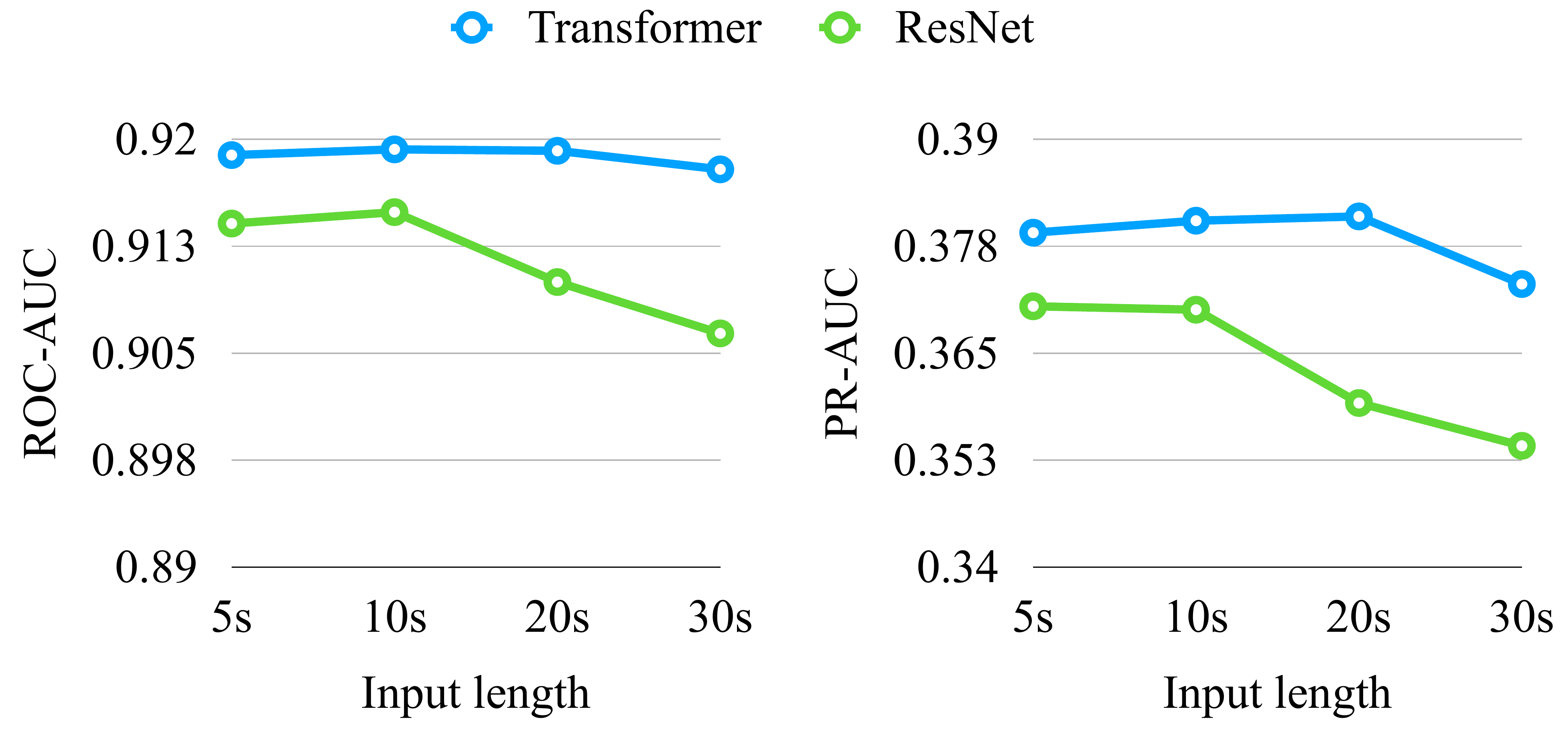}}
 \caption{Performance with different input lengths.}
 \label{fig:length}
\end{figure}

\begin{table}[t!]
\centering
\begin{tabular}{@{}cccc@{}}
\toprule
width & depth & ROC-AUC & PR-AUC \\ \midrule


32 & 4 & 0.9118 & 0.3528 \\
64 & 4 & 0.9178 & 0.3754 \\
128 & 4 & \textbf{0.9194} & 0.3776 \\
256 & 4 & 0.9191 & \textbf{0.3845} \\
512 & 4 & 0.9177 & 0.3788 \\
768 & 4 & 0.9162 & 0.3707 \\
1024 & 4 & 0.9174 & 0.3736 \\ \midrule

256 & 1 & 0.9180 & 0.3736 \\
256 & 2 & 0.9193 & 0.3805 \\
256 & 4 & \textbf{0.9199} & 0.3814 \\
256 & 8 & 0.9181 & \textbf{0.3826} \\
256 & 12 & 0.9165 & 0.3780 \\ 
256 & 16 & 0.9169 & 0.3785 \\ 

\bottomrule
\end{tabular}
\caption{The performance of Music Tagging Transformer with varying width and depth of the attention layers.}
\label{tab:parameter}
\end{table}

\subsection{More Hyperparameter Search}\label{sec:ablation}
In this section, we present the experiment results with various model configurations.
First, we trained our Music Tagging Transformer and short-chunk ResNet with varying input lengths to assess our proposed model's ability to handle long sequences. As shown in Figure~\ref{fig:length}, on both of the metrics, short-chunk ResNet shows a noticeable performance degradation as the audio input gets longer. This shows that the global max pooling in the short chunk ResNet is not perfectly suitable for a long signal. Meanwhile, the Music Tagging Transformer shows consistent performances in general. An exception is when the input is 30-second long. We suspect the performance drop of the Music Tagging Transformer happens because the model cannot take advantage of random cropping data augmentation effect since the 30-second is the full length of the MSD previews.

Second, we investigate different Transformer parameters to figure out the best performing setup. As summarized in Table~\ref{tab:parameter}, Transformer achieved the best performance with attention channels (width) at 128 and 256, and their depth of 4 and 8 layers. However, these optimal parameters are dataset-dependent; as generally observed, a larger network structure would perform better if a larger amount of training data is provided.




\section{Conclusion}\label{sec:conclusion}
In this paper, we proposed a new architecture, Music Tagging Transformer, and improved its tagging performance with a semi-supervised scheme: noisy student training. Experimental results showed that the proposed architecture outperforms the previous state-of-the-art models in supervised music tagging using the MSD~\cite{bertin2011million}. The results also indicate that the tagging models can be further enhanced using noisy student training -- with either knowledge expansion and knowledge distillation. We also provided an analysis result that shows Music Tagging Transformer can handle long audio inputs better than the previous CNN architectures do.

In future work, our Transformer can be further utilized in various MIR tasks. Since Transformer can perform both sequence-level and token-level classification, it can be used in not only music tagging but also tasks such as beat detection and melody extraction. Finally, by combining the multiple MIR tasks in a multi-task learning scheme, Transformer can be trained as a general purpose music representation learning model.

Another important direction to explore is self-supervised learning. Our Music Tagging Transformer can be pre-trained in a self-supervised scheme such as masked embedding prediction~\cite{devlin2018bert} or contrastive learning~\cite{chen2020simple,grill2020bootstrap,spijkervet2021contrastive}. Finally, the pre-trained model can be optimized together with semi-supervised learning to further improve the performance~\cite{chen2020big}.

\section{Acknowledgement}\label{sec:acknowledgement}
This work was funded by the predoctoral grant MDM-2015-0502-17-2 from the Spanish Ministry of Economy and Competitiveness linked to the Maria de Maeztu Units of Excellence Programme (MDM-2015-0502).



\bibliography{ISMIRtemplate}

\end{document}